\documentclass{aa}

\usepackage{graphicx}
\usepackage{txfonts}
\usepackage{xcolor}
\usepackage{multirow}
\usepackage{tabularx}
\usepackage{subfigure}
\usepackage{amsmath}
\usepackage{etoolbox}
\usepackage{microtype}
\usepackage{placeins}
\usepackage{dblfloatfix}
\usepackage{float}

\defcitealias{wil82}{WF82}

\begin{document}

\flushbottom

\title{A homogeneous analysis of archival and new low-resolution spectra of YZ~Cancri}

\author{Zhibin Dai\inst{1,2,3}\corrauth{zhibin\_dai@ynao.ac.cn}
\and Lihuan Yu\inst{1,2,3}
\and Jiao Li\inst{1,2,3}
\and Xuefei Chen\inst{1,2,3}
\and Zhanwen Han\inst{1,2,3}}

\institute{Yunnan Observatories, Chinese Academy of Sciences, 396 Yangfangwang, Guandu District, Kunming 650216, P. R. China
\and International Centre of Supernovae (ICESUN), Yunnan Key Laboratory of Supernova Research, Yunnan Observatories, CAS, Kunming 650216, China
\and Key Laboratory for the Structure and Evolution of Celestial Objects, CAS, Kunming 650216, China}

\date{Received date / Accepted date}

\abstract
{YZ Cancri is an SU UMa-type dwarf nova for which the available optical spectroscopy is sparse and heterogeneous. We present a homogeneous analysis of archival outburst-related spectra and seven new low-resolution spectra obtained during quiescence.}
{We aim to consistently characterize the Balmer, \ion{He}{I}, \ion{Fe}{II}, \ion{O}{I}, and occasional \ion{He}{II} and Bowen features in YZ~Cnc and to identify which state-dependent differences are robustly supported by the available data.}
{We applied a common continuum-normalization and line-measurement procedure to the calibrated spectra. We measured equivalent widths, integrated line intensities, and selected profile parameters, and compared line morphologies and intensity ratios between quiescent and outburst-related states. Because the spectra have a modest and nonuniform resolving power and are not orbital-phase resolved, the profile widths and centroid shifts were used only as relative diagnostics.}
{The calibrated outburst-related spectra are dominated by broad Balmer absorption or composite profiles, whereas the quiescent spectra show strong single-peaked Balmer and \ion{He}{I} emission together with recurrent \ion{Fe}{II} $\lambda$5169 and weak \ion{O}{I} $\lambda$7772 emission. \ion{He}{II} $\lambda$4686 is detected in only two quiescent exposures, and the Bowen blend in one. The quiescent Balmer decrement is relatively flat, while the \ion{He}{I} ratios show substantial exposure-to-exposure scatter. The measured \ion{Fe}{II} ratios show no clear systematic departure from the adopted optically thin reference values, although blending limits their interpretation.}
{The spectra establish an empirical state-dependent description of YZ Cnc. The results are consistent with a nonuniform dwarf-nova accretion environment, but the present low-resolution, non-phase-resolved data do not uniquely determine the spatial origin or geometry of the line-forming gas. The homogeneous measurements provide a reference dataset for future phase-resolved, higher-resolution spectroscopy.}


\maketitle
\nolinenumbers

\section{Introduction}
\label{sec:Introduction}
Cataclysmic variables are interacting binaries in which a white dwarf accretes matter from a Roche-lobe-filling companion. In dwarf novae, the accretion disk alternates between quiescence and outburst through the thermal-instability cycle \citep{war95}. Their optical spectra change accordingly: quiescent spectra are commonly dominated by Balmer and \ion{He}{I} emission, whereas outburst spectra may show broad absorption troughs with narrow emission cores superposed on them, as reported by \citeauthor{wil82} (\citeyear{wil82}; hereafter \citetalias{wil82}) and several subsequent studies \citep{hor86,mar87,woo90,woo92}.

Optical line strengths, profiles, and intensity ratios provide empirical diagnostics of the line-forming gas, although their interpretation depends on spectral resolution, blending, optical depth, and temporal coverage. Balmer and \ion{He}{I} lines trace the dominant optical emission, while recurrent \ion{Fe}{II}, \ion{O}{I}, \ion{He}{II}, and Bowen features can reveal additional low- or high-excitation contributions \citep{sch89,mas00,mas05,pap08,neu16}. In low-resolution spectra, these diagnostics are best used for relative comparisons rather than for uniquely locating the emitting gas.

YZ~Cancri (YZ~Cnc) is an SU~UMa-type dwarf nova near the lower edge of the period gap of cataclysmic variables. Its orbital period is close to two hours, with published values of $P_{\rm orb}=0.0868(2)$~d and $0.086924(7)$~d, and its inclination is estimated to be approximately $25^{\circ}$--$43^{\circ}$ \citep{sha88,van94,zha05}. Studies of other low-inclination dwarf novae have shown that their optical variability and emission structures may require non-axisymmetric or vertically nonuniform accretion disks, including edge or surface hotspot components \citep{dai18,dai20,dai21}. Previous ultraviolet observations revealed phase-dependent resonance-line changes and possible outflow signatures \citep{dre88,woo92}, while X-ray studies showed accretion-state-dependent high-energy emission \citep{cor84,van87,era91,ver97,ver99,hak04}. However, the available optical spectroscopy of YZ~Cnc remains sparse, consisting of published observations and public database spectra \citep{sha88,kaf21,baa_specdb}. The archival spectra analyzed here have not previously been measured together with the new quiescent spectra using a common procedure.

In this work we analyze calibrated archival low-resolution spectra obtained during outburst-related states together with seven new quiescent spectra. The data cover the Balmer series from H$\epsilon$ to H$\alpha$, several \ion{He}{I} transitions, the \ion{Fe}{II} (42) triplet, weak \ion{O}{I}~$\lambda7772$, and occasional \ion{He}{II}~$\lambda$4686 and \ion{C}{III} and \ion{N}{III} Bowen blend emission. We apply a common measurement procedure to quantify the line strengths, selected intensity ratios, and profile properties.

Our primary aim is to establish a homogeneous empirical description of the optical line spectrum of YZ~Cnc and its dependence on luminosity state. We examine which spectral differences are robust across the available observations and compare them with the general phenomenology of dwarf-nova accretion disks. Because the data are low resolution, sparsely sampled, and not orbital-phase resolved, we do not attempt to derive a unique spatial geometry or detailed velocity field for the line-forming gas.

The paper is organized as follows. Section~\ref{sec:Observations_and_data_reduction} describes the spectra and their reduction. Section~\ref{sec:Results} presents the measurement procedure and the empirical line properties. Section~\ref{sec:Discussion_and_Conclusions} discusses the state-dependent behavior, its limitations, and the main conclusions.

\begin{table*}
\caption{Optical spectra used in this work.}\label{tab:spectroscopy}
\centering
\begin{tabular}{lccccccc}
\hline\hline
ID &
UT date &
JD$_{\rm start}$ &
Coverage &
Exposure &
Resolution &
State &
Use \\
\hline
&& 2400000+ & \AA & s &&\\
\hline
B1 & 2018-04-09 & 58217.6911 & 3805--7391 & $1\times2739$ & reported $\simeq874$ & Quiescence & Qual. \\
B2 & 2018-04-11 & 58219.6946 & 3776--7385 & $1\times2510$ & reported $\simeq939$ & Rising & Qual. \\
A1 & 2020-01-03 & 58851.8946 & 3790--7100 & $1\times3146$ & reported $\simeq519$ & Fast decline & Quant. \\
B4 & 2020-01-06 & 58854.9431 & 6260--8900 & $1\times2860$ & reported $\simeq822$ & Post-decline quiescence & Qual. \\
A2 & 2021-04-05 & 59310.4387 & 3785--7230 & $1\times4581$ & reported $\simeq564$ & Normal-outburst peak & Quant. \\
A3 & 2021-04-15 & 59320.3462 & 3650--7400 & $1\times5812$ & reported $\simeq600$ & Pre-superoutburst & Quant. \\
G1 & 2023-05-13 & 60078.0319 & 3830--8350 & $2\times600+1\times900$ & measured $250$--$420$ & Quiescence & Quant. \\
G2 & 2023-05-14 & 60079.0373 & 3830--8350 & $2\times600$ & measured $280$--$430$ & Quiescence & Quant. \\
G3 & 2023-05-14 & 60079.0515 & 3830--8350 & $2\times900$ & measured $280$--$430$ & Quiescence & Quant. \\
B7 & 2025-12-20 & 61030.0074 & 3665--7400 & $1\times6174$ & reported $\simeq476$ & Quiescence & Qual. \\
\hline
\end{tabular}
\tablefoot{JD$_{\rm start}$ gives the start time of the first exposure in each listed spectrum or observing sequence. Barycentric corrections were calculated separately for each individual exposure using its own mid-exposure time derived from the FITS timing information and exposure duration. The spectra are listed in chronological order. The identifiers A1, A2, and A3 refer to the AAVSO versions of the spectra also listed in the BAA database as B3, B5, and B6, respectively. Only A1--A3 and G1--G3 are used for quantitative line measurements (``Quant.''). B1, B2, B4, and B7 lack complete flux calibration, flat-field calibration, or both, and are used only for qualitative comparison (``Qual.''). The resolving powers of the archival spectra are adopted from the corresponding database records and literature descriptions. For the BFOSC spectra (G1--G3), the effective resolving power was measured directly from FeAr lamp profiles obtained with the same instrumental setup. The wavelength-dependent instrumental FWHM were converted into effective resolving powers through $R=\lambda/{\rm FWHM}_{\rm inst}(\lambda)$ and were used for all instrumental-broadening corrections. The wavelength scales of all quantitatively analyzed spectra were independently corrected to the Solar System barycentric frame. The applied corrections were $+8.49$, $-28.86$, and $-29.49~{\rm km\,s^{-1}}$ for A1, A2, and A3, respectively, and ranged from $-27.23$ to $-27.04~{\rm km\,s^{-1}}$ for the seven individual BFOSC exposures.}
\end{table*}

\section{Observations and data reduction}
\label{sec:Observations_and_data_reduction}
Table~\ref{tab:spectroscopy} summarizes the spectra analyzed in this work. We combine public low-resolution spectra from the British Astronomical Association (BAA) Spectroscopy Database \citep{baa_specdb} and the American Association of Variable Star Observers (AAVSO) Spectroscopic Database \citep{kaf21} with seven follow-up spectra obtained in quiescence with the Beijing Faint Object Spectrograph and Camera (BFOSC). The luminosity-state assignments in Table~\ref{tab:spectroscopy} were based on the contemporaneous or nearest available AAVSO photometry. They are used to provide the broad observational context of each spectrum rather than to define a uniformly sampled outburst sequence. Because the archival spectra are isolated exposures and the photometric sampling is not identical for every date, the state labels should be interpreted as approximate state classifications. The dataset includes calibrated spectra obtained during three outburst-related stages and seven new quiescent exposures.

\subsection{Archival spectra}
\label{sec:archival_spectra}
The archival spectra cover 2018 April 9--2025 December 20, with wavelength ranges of 3650--8900~\AA\ and resolving powers of $R\simeq476$--939, depending on the individual exposure. The BAA entries B3, B5, and B6 duplicate the AAVSO spectra A1, A2, and A3, respectively. To avoid ambiguous dual identifiers, we use the AAVSO labels A1, A2, and A3 throughout this paper. The corresponding BAA identifiers are given only in the note to Table~\ref{tab:spectroscopy}. The calibrated A1--A3 spectra are used for the quantitative outburst-related comparison. A1, A2, and A3 correspond to the fast-decline (FD), normal-outburst-peak, and pre-superoutburst stages, respectively. We applied the same continuum normalization and line-measurement procedures as for our ground-based spectra to keep the measurements internally consistent.

For the archival AAVSO and BAA spectra, we retained the wavelength solutions supplied with the database products and did not attempt an independent comparison-lamp recalibration. However, the BAA Flexible Image Transport System (FITS) files associated with the quantitative spectra A1, A2, and A3 contain the observation times, exposure durations, target coordinates, and observatory information required to calculate the barycentric correction for each exposure. Their wavelength scales were therefore shifted to the Solar System barycentric frame using the mid-exposure time of each spectrum before the centroid velocities were measured.

Since no contemporaneous comparison-lamp spectra are available for these public data, the instrumental full width at half maximum (FWHM) was estimated from the resolving power reported for each spectrum, adopting ${\rm FWHM}_{\rm inst}(\lambda)\simeq\lambda/R$ for each measured line. These resolving-power estimates were used only for the instrumental-broadening correction of the line widths and not to redefine the original database wavelength solutions.

\begin{figure*}[!t]
        \sidecaption
    \begin{minipage}{12cm}
    \centering
    \includegraphics[width=\linewidth]{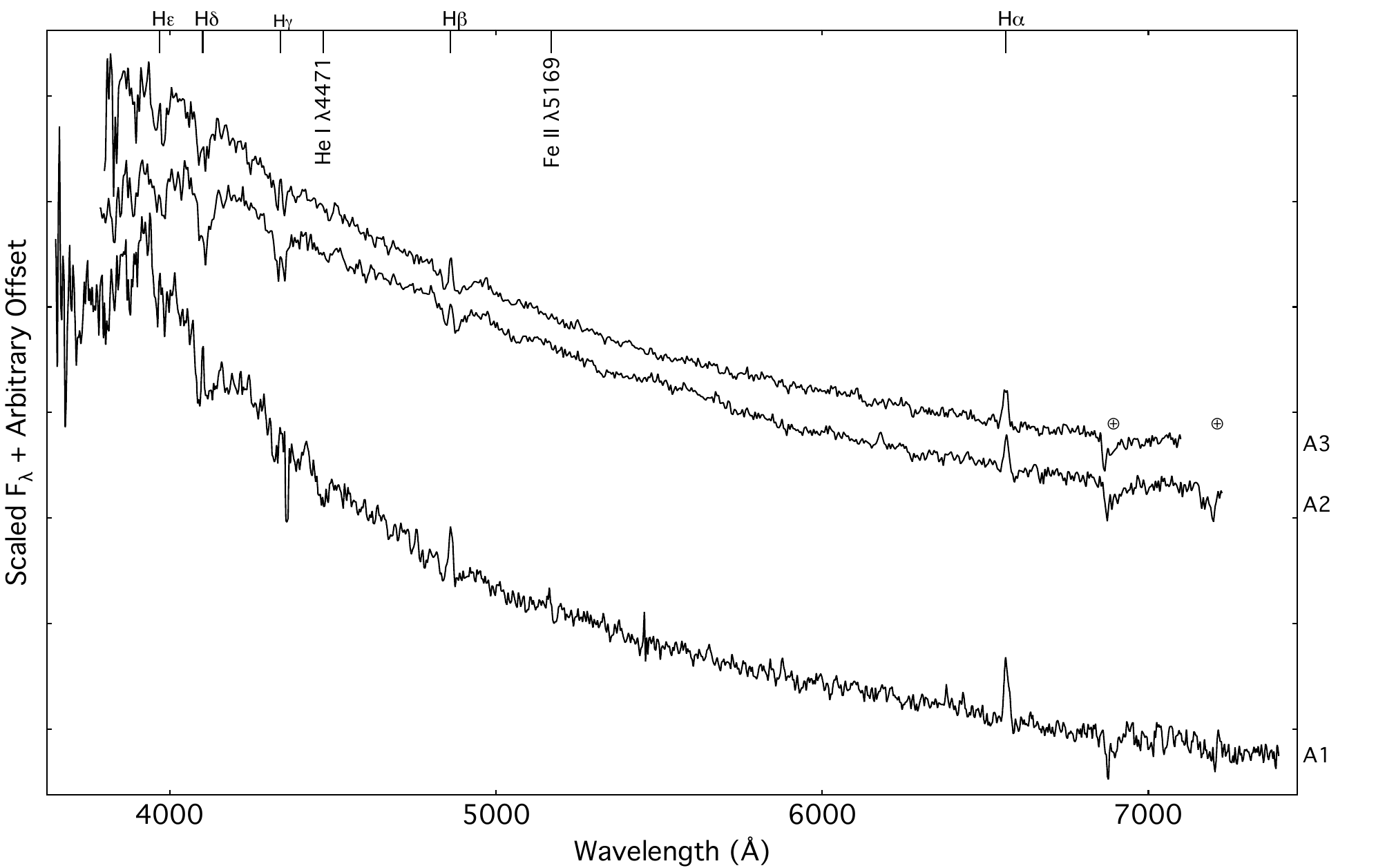}\\[0.5ex]
    \includegraphics[width=\linewidth]{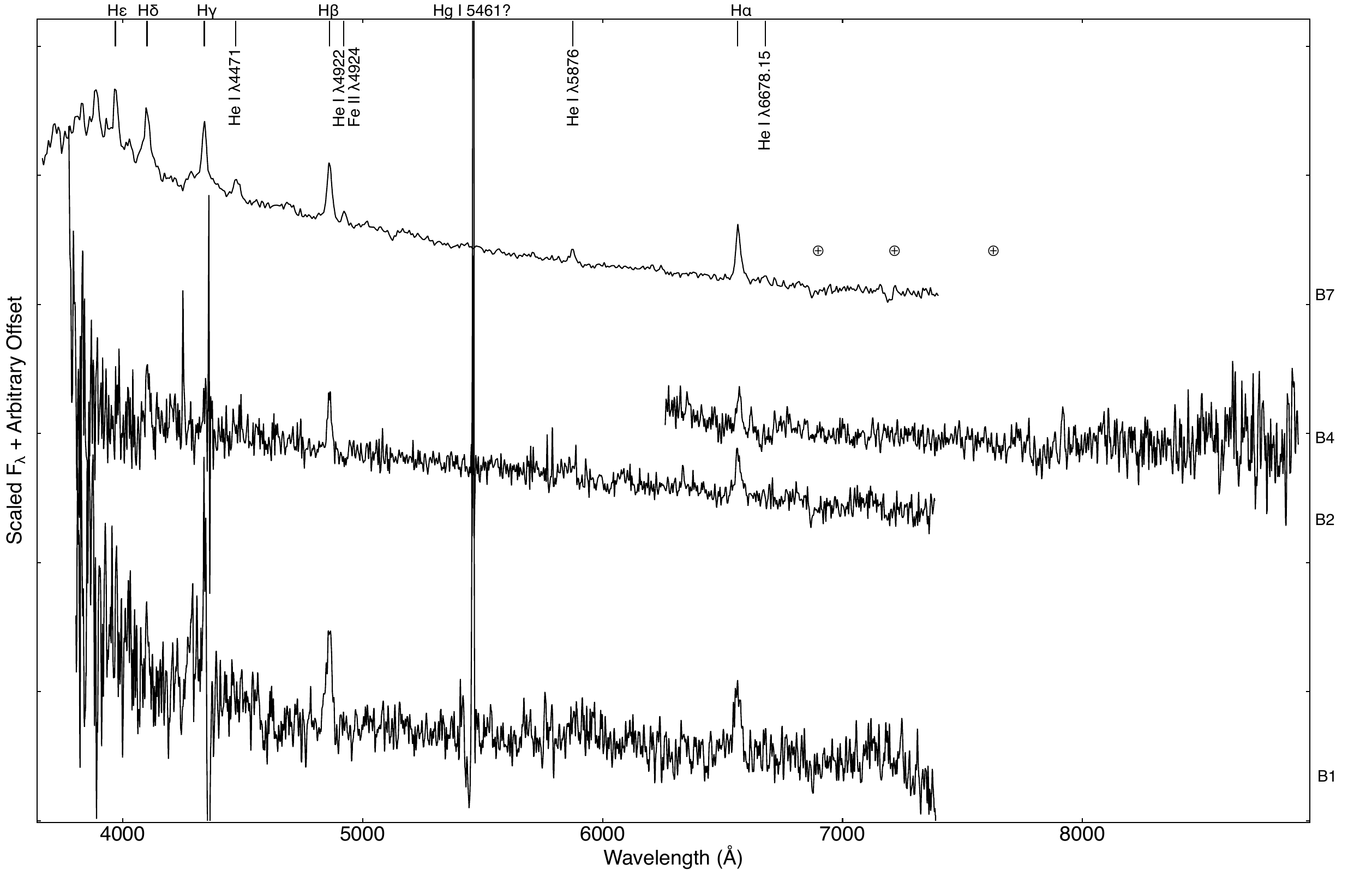}
    \end{minipage}
    \caption{Archival optical spectra of YZ~Cnc obtained from the AAVSO and BAA databases. \textit{Top}: Flux-calibrated spectra A1, A2, and A3 used in the quantitative analysis. \textit{Bottom}: Additional BAA spectra B1, B2, B4, and B7 used only for qualitative comparison. The spectrum identifiers follow Table~\ref{tab:spectroscopy}. For display purposes, spectra are vertically shifted and scaled; these transformations were not used in the quantitative measurements. Prominent features of \ion{H}{I}, \ion{He}{I}, and \ion{Fe}{II} are marked for reference, and the $\oplus$ symbols indicate the principal telluric absorption regions. A narrow feature near $\lambda$5461~\AA\ in B1 and B2 is marked as a possible \ion{Hg}{I} $\lambda5460.7$~\AA\ sky residual. The absence of the original two-dimensional data prevents a definitive assessment, and the feature is excluded from all quantitative analyses.}
        \label{fig:aavso_spectra}
\end{figure*}

The remaining BAA spectra, B1, B2, B4, and B7, lack flux calibration, flat-field calibration, or both. They are therefore used only for qualitative reference in Fig.~\ref{fig:aavso_spectra}, not for line-intensity ratios, equivalent-width (EW) measurements, or other quantitative diagnostics. A narrow feature near $\lambda$5461~\AA\ appears in the B1 and B2 spectra. Its wavelength is consistent with the \ion{Hg}{I} $\lambda$5460.7~\AA\ night-sky line commonly associated with artificial-light contamination. Because the original two-dimensional frames and sky spectra are unavailable, we cannot verify the subtraction directly. We therefore regard the feature as a probable sky residual or reduction artifact and exclude it from all astrophysical interpretation and quantitative measurements. We also searched LAMOST DR11, SDSS SkyServer, DESI DR1, and \textit{Gaia} DR3 for additional optical spectra. No product with sufficient spectral sampling and calibration for the present line measurements was found. In particular, the \textit{Gaia} DR3 BP and RP spectrum\footnote{Source$\rm _{id}$ 683908812437792000 \citep{gai23,dea23} contains only 41 sampled points over 400--800~nm with a 10~nm step, corresponding to $R_{\rm samp}\simeq$40--80.} is too coarsely sampled and was excluded from the analysis.

Several published orbital ephemerides exist for YZ~Cnc, but extrapolating them to the 2020--2025 observations introduces phase uncertainties exceeding one orbital cycle. We therefore do not assign absolute orbital phases and treat centroid shifts only as single-epoch profile diagnostics.

\subsection{Ground-based spectra}
\label{sec:ground-based_spectra}
We obtained seven quiescent spectra of YZ~Cnc on 2023 May 13--14 with the BFOSC on the Xinglong 2.16~m telescope. The observations used grating G4, giving a dispersion of 2.97~\AA\,pixel$^{-1}$ and a wavelength coverage of about 3830--8350~\AA. The G1 sequence was obtained on May 13 and consists of two 600~s exposures and one 900~s exposure; the final integration was extended to 900~s in response to changing observing conditions. The G2 and G3 sequences were obtained on May 14 and consist of two 600~s and two 900~s exposures, respectively.

The BFOSC spectra were reduced with standard long-slit procedures in the Image Reduction and Analysis Facility (IRAF; \citealt{tod86,tod93}), including overscan and bias subtraction, flat-field correction, one-dimensional extraction, wavelength calibration with FeAr comparison-lamp spectra, and flux calibration. After the FeAr wavelength calibration, each BFOSC exposure was independently shifted to the Solar System barycentric frame using the target coordinates, the geographic location of the Xinglong Observatory, and the corresponding mid-exposure time. Three spectrophotometric standard stars\footnote{BD+33$^\circ$2642, HD~74721, and HD~161817} were observed, and HD~161817 was used as the common sensitivity reference for the two observing nights. Because exposure-dependent slit losses could not be independently corrected, absolute line intensities, $I_{\rm Line}$, are treated cautiously and line ratios within individual spectra are emphasized.

The same FeAr lamp spectra were also used to estimate the instrumental FWHM as a function of wavelength. Isolated and reliably fit FeAr lines were first measured with Gaussian profiles. For each science line, ${\rm FWHM}_{\rm inst}$ was then assigned from nearby reliable lamp lines using a local weighted estimate; when local lamp lines were sparse, a smooth empirical ${\rm FWHM}_{\rm inst}(\lambda)$ relation constructed from the reliable FeAr lines was used. This procedure avoids applying a single constant instrumental FWHM to all BFOSC lines across the full wavelength range.

\begin{figure*}[!t]
        \centering
        \includegraphics[width=\textwidth,height=0.38\textheight]{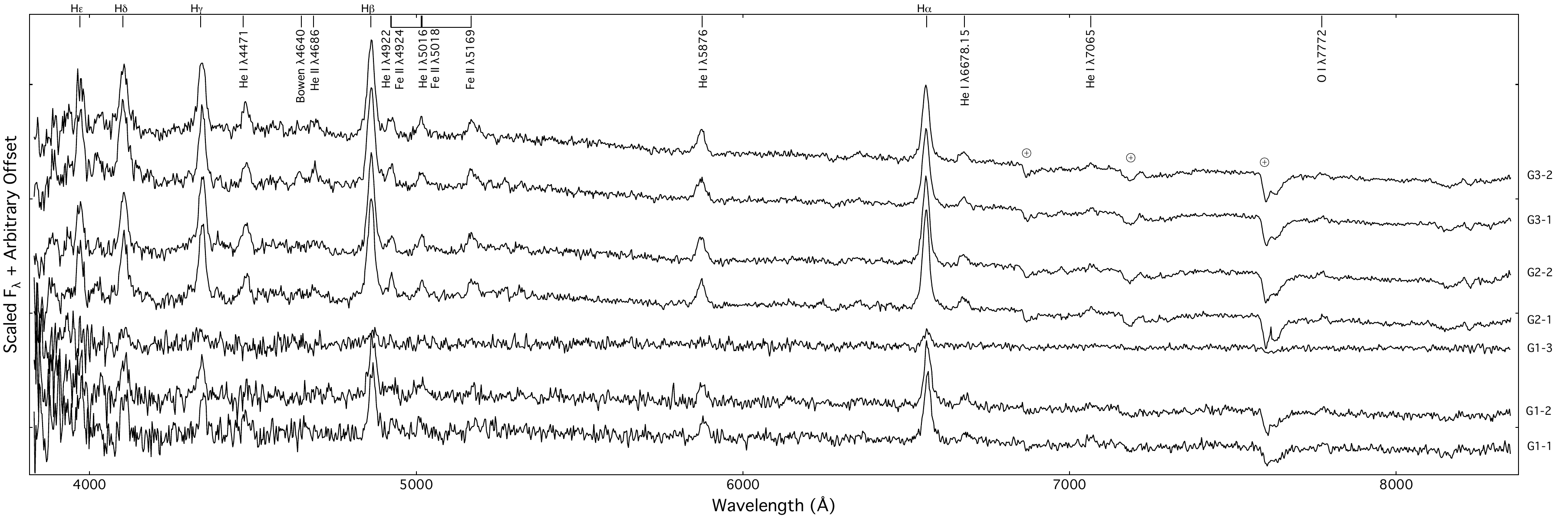}
    \caption{Seven flux-calibrated optical spectra of YZ~Cnc obtained with the BFOSC instrument on the Xinglong 2.16~m telescope. The spectra cover the wavelength range of approximately 3830--8350~\AA. The labels G1-1 through G3-2 identify the individual exposures within the G1-G3 sequences listed in Table~\ref{tab:spectroscopy}. Emission features of \ion{H}{I}, \ion{He}{I}, \ion{Fe}{II}, \ion{O}{I}, and occasional \ion{He}{II} are marked for identification. The Bowen blend near $\lambda$4640--4650~\AA\ is marked as a combined \ion{C}{III} and \ion{N}{III} feature. The $\oplus$ symbols mark the principal telluric absorption regions. For display purposes, spectra were scaled and vertically shifted by arbitrary offsets. All EWs, $I_{\rm Line}$s, velocities, and profile measurements were performed on the original flux-calibrated spectra before these transformations.}
        \label{fig:ground_spectra}
\end{figure*}

For display in Fig.~\ref{fig:ground_spectra}, the flux-calibrated spectra were multiplied by individual constant factors and shifted vertically by arbitrary additive offsets. As shown in Fig.~\ref{fig:ground_spectra}, the quiescent spectra display strong single-peaked Balmer emission from H$\epsilon$ to H$\alpha$, multiple \ion{He}{I} triplet and singlet lines, the \ion{Fe}{II} (42) triplet, weak \ion{O}{I} $\lambda$7772, and occasional \ion{He}{II} $\lambda$4686 plus \ion{C}{III} and \ion{N}{III} Bowen blend emission.

\subsection{Dereddening}
\label{sec:dereddening}
YZ~Cnc is nearby and lies at relatively high Galactic latitude (i.e., $b\simeq+28.8^{\circ}$), so foreground reddening is expected to be negligible. 
We adopted the \textit{Gaia} EDR3 geometric distance estimate $d_{50}=233.49$~pc, with $d_{16}=231.68$~pc and $d_{84}=235.03$~pc, from \citet{bai21}, and queried the Bayestar19 three-dimensional dust map \citep{gre19} at these distances. The map returned $E(B-V)=0.0$ at the adopted distance estimates. Therefore no extinction correction was applied. This choice has no practical effect on the optical line ratios used here; any unmodeled residual reddening would mainly affect the bluest Balmer lines. Given the negligible extinction estimate, we do not propagate reddening uncertainties into line ratios.

\section{Results}
\label{sec:Results}
\subsection{Line-measurement procedure}
\label{sec:Line-measurement_procedure}
All line measurements were performed with a custom Python-based procedure applied consistently to the calibrated archival and BFOSC spectra. The same continuum-normalization, profile-fitting, and uncertainty-estimation scheme was used for the outburst-related composite profiles and the quiescent emission features. All measurements were made on the original calibrated spectra, before the multiplicative scaling and additive offsets used only for display in Figs.~\ref{fig:aavso_spectra} and \ref{fig:ground_spectra}.

For continuum normalization, a Savitzky--Golay-smoothed spectrum \citep{sav64} was used to identify emission and absorption features, which were masked before fitting a smoothing spline to the remaining line-free regions. The continuum was determined independently for each spectrum, and the original spectrum was divided by this spline for EW and profile measurements. Pure emission and absorption profiles are denoted by E and A, respectively. Isolated lines were measured with single- or double-Gaussian fits over several nearby fitting windows, and the accepted solutions were combined using robust central values and the scatter among the trials. All fits were visually inspected, with uncertain cases remeasured interactively.

The observed Gaussian widths were corrected for instrumental broadening as
\begin{equation}
{\rm FWHM}_{\rm corr} = \left({\rm FWHM}_{\rm obs}^{2} - {\rm FWHM}_{\rm inst}^{2}\right)^{1/2},
\end{equation}
where ${\rm FWHM}_{\rm inst}$ was assigned separately for each transition from the wavelength-dependent FeAr measurements for the BFOSC spectra or from the reported resolving power for the archival spectra. Uncertainties in ${\rm FWHM}_{\rm corr}$ were propagated from the observed and instrumental widths. No corrected width was assigned when ${\rm FWHM}_{\rm obs}\leq{\rm FWHM}_{\rm inst}$; measurements with $1<{\rm FWHM}_{\rm obs}/{\rm FWHM}_{\rm inst}<1.3$ were flagged as marginally resolved. Line centroids were measured relative to the adopted laboratory wavelengths on the barycentric-corrected wavelength scales described in Sect.~\ref{sec:Observations_and_data_reduction}.

For the composite outburst-related profiles, the broad background and narrow central component were separated phenomenologically. Four interactively selected boundary points defined the full profile and central-core intervals. After masking the core, the broad component was represented by a low-order polynomial fit to the two side regions, and the central component was measured relative to this local background. The labels ``B,'' ``C,'' and ``T'' denote the broad background, narrow central component, and total measured profile, respectively.

The EWs were measured from the locally continuum-normalized spectra, whereas the $I_{\rm Line}$ values were obtained from the corresponding continuum-subtracted calibrated spectra. H$\beta$ was adopted as the reference transition for the Balmer ratios. With the sign convention used in the electronic Tables~5 and 6 at the CDS, emission has a negative EW and a positive $I_{\rm Line}$, while absorption has a positive EW and a negative $I_{\rm Line}$.

Measurement uncertainties were estimated from 1000 Monte Carlo noise realizations based on the residual scatter around the adopted continuum and were combined with the scatter among accepted fitting windows. Because the continuum was held fixed in the simulations, the quoted uncertainties represent statistical errors conditional on the adopted continuum placement. The complete archival and BFOSC line measurements are provided in electronic Tables~5 and 6, respectively, at the CDS.

At the present resolution, \ion{Fe}{II} $\lambda$4924 and $\lambda$5018 remain blended with \ion{He}{I} $\lambda$4922 and $\lambda$5016, respectively, whereas $\lambda$5169 is the cleanest member of the \ion{Fe}{II} multiplet-42 triplet. The feature near 7772--7775~\AA\ is treated as the unresolved \ion{O}{I} triplet, and the feature around $\lambda$4640--4650~\AA\ as the unresolved \ion{C}{III} and \ion{N}{III} Bowen blend.

\subsection{State-dependent line morphology and strengths}
\label{sec:state_dependent_results}
The calibrated spectra show a clear state-dependent change in line morphology. The outburst-related spectra A2 and A3 are dominated by broad Balmer absorption profiles with weaker central emission components, while the FD spectrum A1 contains more complex composite profiles, including absorption-like \ion{He}{I}~$\lambda$4471 and a composite \ion{Fe}{II}~$\lambda$5169 feature. Because each outburst-related stage is represented by only one calibrated spectrum, these differences are treated as descriptive rather than as fully sampled evolutionary trends. This overall change from absorption and composite profiles in the outburst-related spectra to emission-dominated quiescent spectra is typical of dwarf novae \citep{wil82,mar87,woo90,szk18}; the YZ~Cnc-specific information here lies in the recurrence and intermittency of the weaker features described below.

The seven BFOSC quiescent spectra (Fig.~\ref{fig:ground_spectra}) are emission-dominated. Balmer emission from H$\epsilon$ to H$\alpha$ is present throughout the sequence, together with several \ion{He}{I} lines. \ion{Fe}{II}~$\lambda$5169, the least blended member of multiplet~42, is detected in five quiescent exposures. Weak \ion{O}{I}~$\lambda$7772 is present in most exposures, while \ion{He}{II}~$\lambda$4686 is detected only in G3-1 and G3-2 and the Bowen blend only in G3-1. The occurrence of these features is summarized in Table~\ref{tab:key_diagnostics}, while selected Balmer and ionic line ratios are listed in Tables~\ref{tab:key_Balmer_ratios} and \ref{tab:key_ratios}.

The EW measurements quantify the same contrast. Balmer lines have the largest absolute quiescent EWs, \ion{He}{I} lines are weaker but recurrent, and the \ion{Fe}{II} and \ion{O}{I} features occupy a still weaker range. The high-excitation features are both weak and intermittent.

The measured profile widths have substantial uncertainties, and several features are only marginally resolved. The centroid measurements likewise show no clear systematic velocity offset shared by the different transitions. These measurements are therefore retained in the electronic Tables~5 and 6 at the CDS as part of the homogeneous line dataset, but the present resolution and sampling do not provide additional kinematic constraints beyond the observed line morphology.

\begin{table*}[!tbp]
\caption{Key spectral profiles in YZ~Cnc.}
\label{tab:key_diagnostics}
\centering
\normalsize
\setlength{\tabcolsep}{4pt}
\begin{tabular}{cccccc}
\hline\hline
Spectrum & Balmer & \ion{He}{I} & \ion{Fe}{II} & \ion{O}{I} $\lambda$7772 & \ion{He}{II} and Bowen \\
\hline
A1 & Composite and emission & Absorption at $\lambda$4471 & Composite $\lambda$5169 & Not detected & Not detected \\
A2 & Composite & Not detected & Not detected & Not detected & Not detected \\
A3 & Composite & Not detected & Not detected & Not detected & Not detected \\
G1 & Emission & Some exposures & Some exposures & Emission & Not detected \\
G2 & Emission & Emission & Emission & Emission & Not detected \\
G3 & Emission & Emission & Emission & Emission & Some exposures \\
\hline
\end{tabular}
\tablefoot{Full line measurements and auxiliary profile parameters are provided in electronic Tables~5 and 6 at the CDS. Entries for G1--G3 summarize all exposures in the corresponding sequence; ``some exposures'' indicates that the feature was not detected in every individual exposure.}
\end{table*}
\begin{figure}[!tbp]
        \centering
        \includegraphics[width=0.98\columnwidth,angle=0]{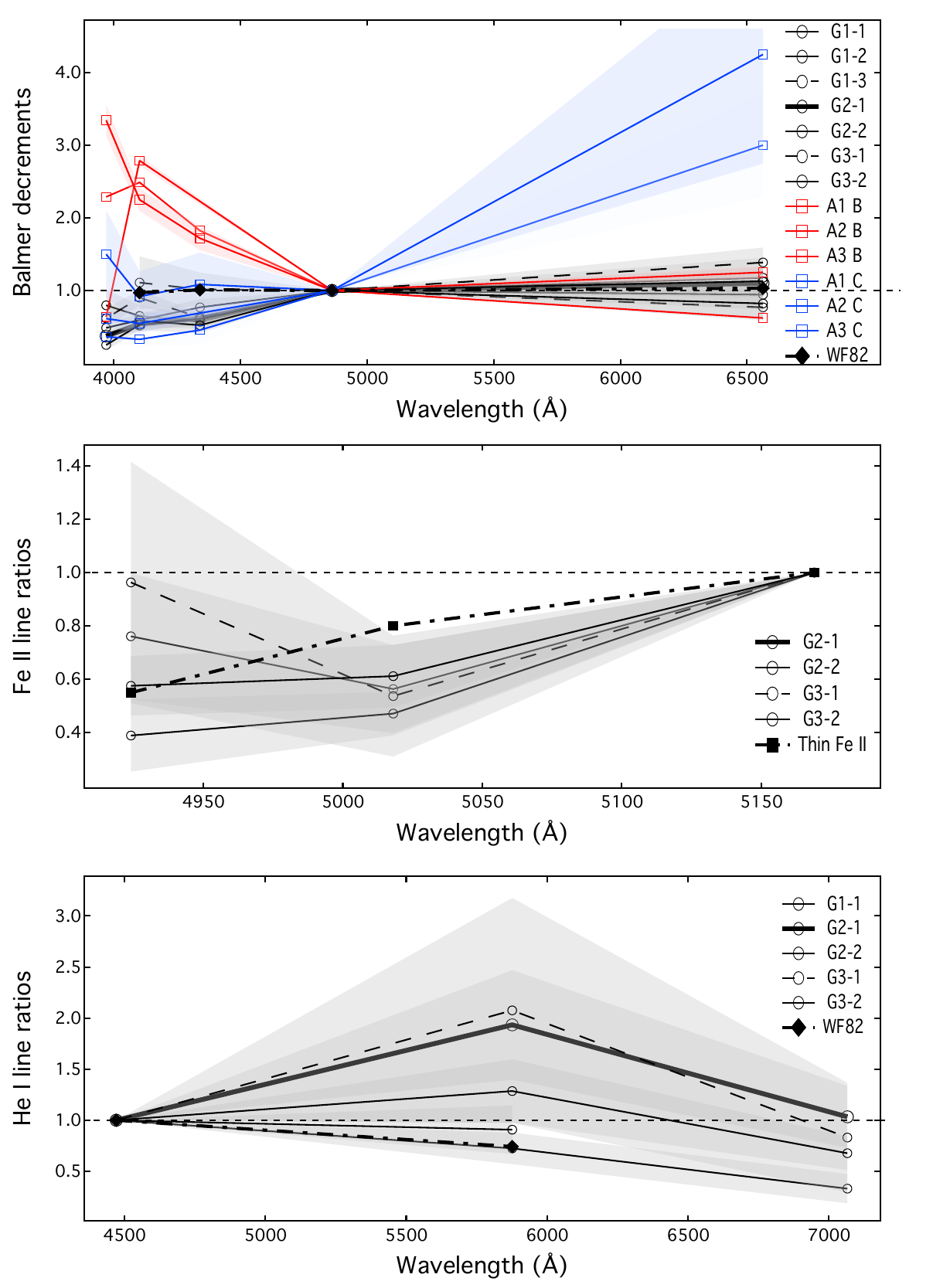}
    \caption{Line-intensity ratios measured from the spectra of YZ~Cnc. \textit{Top}: Balmer decrements normalized to H$\beta$. \textit{Middle}: \ion{Fe}{II} multiplet-42 line ratios ($\lambda$4924, $\lambda$5018, and $\lambda$5169) normalized to \ion{Fe}{II} $\lambda$5169. \textit{Bottom}: \ion{He}{I} triplet line ratios normalized to \ion{He}{I} $\lambda4471$. Symbols identifying the individual BFOSC exposures and the decomposed archival components are given separately in each panel. The \citeauthor{wil82} (\citeyear{wil82}; hereafter \citetalias{wil82}) values provide an empirical dwarf-nova comparison for the Balmer and \ion{He}{I} ratios. The filled black squares in the \ion{Fe}{II} panel show the optically thin reference ratios calculated from the adopted atomic transition data \citep{fuh06,mel09,deb14}. Translucent bands around measured curves show the propagated line-ratio uncertainties. Connecting lines are included only to guide the eye between discrete transitions, and the horizontal dashed line at unity marks the normalization line in each panel. The wavelength ranges are chosen independently for the three panels to display the relevant transitions clearly.}
        \label{fig:decrements}
\end{figure}

\subsection{Line-intensity ratios}
\label{sec:line_ratio_results}
Because exposure-dependent slit losses could not be independently corrected, we compare line ratios measured within individual spectra rather than absolute line-flux differences between exposures. Figure~\ref{fig:decrements} summarizes three complementary line-ratio comparisons. The Balmer panel shows that the quiescent spectra have a relatively flat decrement, with $I_{\rm Line}({\rm H}\alpha)/I_{\rm Line}({\rm H}\beta)$ ranging from approximately 0.8 to 1.4 and generally remaining of order unity. This behavior is flatter than the steeper decrement expected for a simple optically thin Baker--Menzel Case-B recombination spectrum, indicating that the Balmer-emitting gas cannot be described by such a simple nebular-like limit. The decomposed broad-background and narrow-central components of the outburst-related spectra span a wider range of Balmer ratios. The \citetalias{wil82} values are shown as an empirical comparison with the line-ratio behavior reported for dwarf novae.

The \ion{He}{I} ratios show substantially larger exposure-to-exposure scatter than the Balmer ratios and do not define a single reproducible sequence across the available quiescent spectra. The measured \ion{Fe}{II} multiplet-42 ratios show no clear systematic departure from the adopted optically thin reference ratios, although $\lambda4924$ and $\lambda5018$ are affected by blending with nearby \ion{He}{I} lines, and therefore do not provide a unique optical-depth constraint. Thus, the principal results of these comparisons are the relatively flat Balmer decrement and its departure from the simple Case-B limit, the larger variability of the \ion{He}{I} line ratios, and the recurrent presence of \ion{Fe}{II} $\lambda5169$. Given the sparse, non-phase-resolved sampling, some of the exposure-to-exposure ratio variations may reflect orbital-phase variability in addition to luminosity-state differences.

Table~\ref{tab:key_ratios} provides several additional ratios that complement the same-species comparisons in Fig.~\ref{fig:decrements}. The \ion{He}{I}\,$\lambda$6678/5876 ratio compares the singlet and triplet line strengths, while \ion{He}{I}\,$\lambda$5876/H$\beta$ measures the relative strength of helium and Balmer emission. The \ion{Fe}{II}\,$\lambda$5169/H$\beta$ and \ion{He}{II}\,$\lambda$4686/H$\beta$ ratios provide empirical measures of the low- and high-excitation contributions, respectively. Where measurable, \ion{He}{I}\,$\lambda$5876/H$\beta$ remains within approximately 0.19--0.25, while \ion{Fe}{II}\,$\lambda$5169/H$\beta$ spans approximately 0.11--0.21. The \ion{He}{I}\,$\lambda$6678/5876 ratio remains approximately 0.7--1.0 where measurable. \ion{He}{II}\,$\lambda$4686/H$\beta$ is approximately 0.07--0.11 in the two G3 exposures in which \ion{He}{II} is detected. These ratios provide descriptive measures of relative line strength.

\begin{table*}[!tbp]
\caption{Balmer decrements of the decomposed outburst-related profiles.}
\label{tab:key_Balmer_ratios}
\centering
\normalsize
\setlength{\tabcolsep}{4pt}
\begin{tabular}{ccccc}
\hline\hline
Spectrum & $^{a}$Component & $I_{\rm Line}$[H$\alpha$/H$\beta$] & $I_{\rm Line}$[H$\gamma$/H$\beta$] &  $I_{\rm Line}$[H$\delta$/H$\beta$] \\
\hline
A1 & B & \ldots & \ldots & 2.79(0.07) \\
& C & \ldots & \ldots & 0.5(0.1) \\
A2 & B & 0.62(0.04) & 1.82(0.09) & 2.5(0.1) \\
& C & 3.0(0.7) & 0.5(0.2) & 0.3(0.1) \\
A3 & B & 1.25(0.08) & 1.7(0.2) & 2.3(0.2) \\
& C & 4.3(1.5) & 1.1(0.4) & 0.9(0.3) \\
\hline
\end{tabular}
\tablefoot{$^{a}$ For the composite Balmer profiles, the ratios were calculated separately for the fit broad-background (B) and narrow-central (C) components. These component ratios are phenomenological measurements and do not by themselves imply spatially distinct line-forming regions.\\
Values in parentheses are the $1\sigma$ measurement uncertainties.}
\end{table*}
\begin{table*}[!tbp]
\caption{Selected additional quiescent line-intensity ratios.}
\label{tab:key_ratios}
\centering
\normalsize
\setlength{\tabcolsep}{4pt}
\begin{tabular}{ccccc}
\hline\hline
Spectrum & \ion{He}{I}\,$\lambda$6678/5876 & \ion{He}{I}\,$\lambda$5876/H$\beta$ & \ion{Fe}{II}\,$\lambda$5169/H$\beta$ & \ion{He}{II}\,$\lambda$4686/H$\beta$ \\
\hline
G1-1 & 0.7(0.3) & 0.25(0.05) & \ldots & \ldots \\
G1-2 & 1.0(0.3) & 0.23(0.05) & 0.11(0.04) & \ldots \\
G1-3 & \ldots & \ldots & \ldots & \ldots \\
G2-1 & 0.8(0.3) & 0.21(0.03) & 0.18(0.05) & \ldots \\
G2-2 & 0.9(0.3) & 0.19(0.03) & 0.19(0.03) & \ldots \\
G3-1 & 0.8(0.2) & 0.19(0.03) & 0.16(0.06) & 0.11(0.02) \\
G3-2 & 1.0(0.2) & 0.20(0.03) & 0.21(0.03) & 0.07(0.02) \\
\hline
\end{tabular}
\tablefoot{Ratios are calculated from the measured $I_{\rm Line}$ values. No reddening correction was applied because the foreground reddening toward YZ~Cnc is negligible. Ratios involving strongly blended features are omitted. Values in parentheses are the $1\sigma$ measurement uncertainties.}
\end{table*}

\section{Discussion and conclusions}
\label{sec:Discussion_and_Conclusions}
\subsection{State-dependent optical spectrum of YZ~Cnc}
\label{sec:state_dependent_discussion}
The clearest result of the present dataset is the contrast between the sampled outburst-related and quiescent spectra. The calibrated outburst-related spectra show blue continua and broad Balmer absorption or composite profiles, consistent with the familiar appearance of an optically thick dwarf-nova disk during or shortly after outburst \citep{wil82,mar87,woo90,szk18}. The quiescent BFOSC spectra are instead dominated by single-peaked Balmer and \ion{He}{I} emission and contain a richer set of weak metal lines.

This contrast is robust because it is visible directly in the calibrated spectra and does not depend on detailed profile modeling. However, each archival luminosity state is represented by only one quantitative spectrum. The differences among A1, A2, and A3 therefore cannot establish a continuous spectroscopic evolution through the outburst cycle. In particular, the composite \ion{Fe}{II}~$\lambda5169$ profile in A1 should be regarded as a property of that individual FD spectrum rather than as a general signature of all FD stages.

The absence of reliable orbital phases prevents us from separating luminosity-state changes from orbital modulation. Because line strengths and profiles in dwarf novae can vary with orbital phase, orbital modulation is a plausible contributor to some of the observed exposure-to-exposure variations. The present data therefore establish a state-associated empirical contrast, while the relative contributions of luminosity state and orbital phase remain unresolved.

\subsection{Specific line properties in quiescence}
\label{sec:specific_line_properties}

The relatively flat quiescent Balmer decrement is consistent with dense and radiatively processed hydrogen-emitting gas, as commonly observed in quiescent dwarf novae \citep{wil82,mar87,neu02}. It differs from the steeper decrement expected for a tenuous optically thin recombination nebula. The present ratios do not, however, provide a unique temperature, density, or optical-depth solution.

The \ion{He}{I} ratios show substantially greater scatter than the Balmer ratios. Their main observational implication is therefore variability in the relative strengths of the \ion{He}{I} transitions, rather than a single optical-depth sequence. Optical-depth effects, collisional excitation, self-absorption, and continuum-placement uncertainties may all contribute \citep{smi96,ben02,por05}.

A more YZ~Cnc-specific result is the recurrent detection of low-ionization emission in the quiescent spectra. \ion{Fe}{II}~$\lambda$5169 is detected in five BFOSC exposures, and weak \ion{O}{I}~$\lambda$7772 is present in most of them. The other two \ion{Fe}{II} multiplet-42 members are blended with nearby \ion{He}{I} lines, so their ratios cannot uniquely determine the optical depth of the Fe$^{+}$-emitting gas. Nevertheless, the repeated $\lambda$5169 detections establish that low-ionization emission is not confined to a single quiescent exposure.

The high-excitation contribution is more intermittent. \ion{He}{II}~$\lambda$4686 appears only in G3-1 and G3-2, and the Bowen blend is confidently identified only in G3-1. Such features require a harder ionizing field than the dominant Balmer and \ion{He}{I} emission, but the available spectra cannot distinguish among a boundary layer, inner-disk irradiation, stream impact, or short-timescale variability. Their principal significance here is their restricted occurrence within the seven quiescent exposures.

\subsection{Limitations and future observations}
\label{sec:limitations}
The present study is limited by the modest and heterogeneous
spectral resolution, the small number of independent observing
nights and luminosity-state samples, the lack of orbital-phase
coverage, and the incomplete calibration of several archival
spectra. These limitations restrict the present analysis to
empirical comparisons of the observed line properties and
prevent a detailed kinematic interpretation.

The decomposition of the outburst-related composite profiles is also nonunique at the available resolution. The broad-background and narrow-core measurements are useful phenomenological descriptions, but they do not prove that the two components originate in spatially separate regions. Similarly, the simultaneous presence of lines with different ionization requirements is consistent with a nonuniform accretion environment, but does not distinguish discrete layers from a continuous disk atmosphere with gradients.

Future progress requires phase-resolved spectroscopy at a substantially higher spectral resolution. Such observations would test whether the observed line-profile variations and the recurrent \ion{Fe}{II} $\lambda$5169 emission repeat with orbital phase, resolve the \ion{Fe}{II}--\ion{He}{I} blends, and determine whether the intermittent \ion{He}{II} and Bowen emission is tied to preferred orbital phases or to short-timescale accretion variability. Simultaneous photometry would provide an unambiguous luminosity-state and flickering context for each spectrum. Taken together, the available spectra place YZ~Cnc within the usual spectroscopic behavior of dwarf novae, while providing a homogeneous reference dataset for future phase-resolved studies of its recurrent low-ionization and intermittent high-excitation features.

\section*{Data availability}

Tables~5 and 6 are only available in electronic form at the CDS via anonymous ftp to cdsarc.u-strasbg.fr (130.79.128.5) or via \url{http://cdsweb.u-strasbg.fr/cgi-bin/qcat?J/A+A/}.

\begin{acknowledgements}
This work is supported by the National Natural Science Foundation of China (Nos.\ 12288102, 12125303, 12090040, 12090043, 12103064, 12403039, 12373036), the National Key R\&D Program of China (grant Nos. 2021YFA1600403, 2021YFA1600401, 2021YFA1600400), and the Natural Science Foundation of Yunnan Province (Nos. 202201AT070180, 202201BC070003, 202001AW070007), the International Centre of Supernovae, Yunnan Key Laboratory (No. 202302AN360001), the ``Yunnan Revitalization Talent Support Program''--Science, Technology Champion Project (No. 202305AB350003), and Yunnan Fundamental Research Projects (grant Nos. 202501CF070018). JL is supported by the Yunnan Fundamental Research Projects (YFRP) grant No. 202501CF070016 and the Young Talent Project of Yunnan Revitalization Talent Support Program. We thank the staff of the Xinglong 2.16 m telescope for their support during the observations. We also acknowledge with thanks the variable-star observations from the AAVSO International Database and the spectra made available through the BAA Spectroscopy Database, which were used in this work.

The data analysis made use of Astropy (v4.0.2).
\end{acknowledgements}

\end{document}